\documentclass[
superscriptaddress,%groupedaddress,%unsortedaddress,%runinaddress,
nofootinbib, %nobibnotes, %bibnotes,
 amsmath,amssymb,
 aps,prd,
 twocolumn
]{revtex4-2}

\usepackage{graphicx}
\usepackage{bm}
\usepackage{hyperref}
\usepackage{color}
\usepackage[normalem]{ulem}

\makeatletter
\def\maketitle{
\@author@finish
\title@column\titleblock@produce
\suppressfloats[t]}
\makeatother

\newcommand{\beq}{\begin{equation}}
\newcommand{\eeq}{\end{equation}}
\def\bea{\begin{eqnarray}}
\def\eea{\end{eqnarray}}
\def\beal{\begin{align}}
\def\eal{\end{align}}
\newcommand{\bei}{\begin{itemize}}
\newcommand{\eei}{\end{itemize}}
\newcommand{\bmat}{\begin{matrix}}
\newcommand{\emat}{\end{matrix}}

\newcommand{\Fig}[1]{Fig.~\ref{#1}}
\newcommand{\Eq}[1]{Eq.~(\ref{#1})}

\newcommand{\Sec}[1]{Sec.~\ref{#1}}
\newcommand{\App}[1]{App.~\ref{#1}}

\def\={\,=\,}
\def\+{\,+\,}
\def\-{\,-\,}
\def\Tr{{\rm Tr}}
\def\det{{\rm det}}
\def\Mpl{M_{\rm Pl}}
\def\LQCD{\Lambda_{\rm QCD}}
\def\MeV{\,{\rm MeV}}
\def\GeV{\,{\rm GeV}}
\def\vEW{v_{\rm EW}}
\def\QtoV{Q$^2$V\,}

\begin{document}

\title{Hubble selection of the weak scale from QCD quantum critical point}
 
\author{Sunghoon Jung}
 \email{sunghoonj@snu.ac.kr}
 \affiliation{Center for Theoretical Physics, Department of Physics and Astronomy, Seoul National University, Seoul 08826, Korea}
\affiliation{Astronomy Research Center, Seoul National University, Seoul 08826, Korea}

\author{TaeHun Kim}
 \email{gimthcha@snu.ac.kr}
 \affiliation{Center for Theoretical Physics, Department of Physics and Astronomy, Seoul National University, Seoul 08826, Korea}

\begin{abstract}
There is growing evidence that the small weak scale may be related to self-organized criticality.
In this regard, we note that if the strange quark were lighter, the QCD phase transition could have been first order, possibly exhibiting quantum critical points at zero temperature as a function of the Higgs vacuum expectation value $v_h$ smaller than (but near) the weak scale. 
We show that these quantum critical points allow a dynamical selection of the observed weak scale, via quantum-dominated stochastic evolutions of the value of $v_h$ during eternal inflation. Although the values of $v_h$ in different Hubble patches are described by a probability distribution in the multiverse, inflationary quantum dynamics ensures that the peak of the distribution evolves toward critical points (self-organized criticality),  driven mainly by the largest Hubble expansion rate there -- the Hubble selection of the universe. 
To this end, we first explore the quantum critical points of the three-flavor QCD linear sigma model, parametrized by $v_h$ at zero temperature, and we present a relaxion model for the weak scale. Among the patches that have reached reheating, it results in a sharp probability distribution of $v_h$ near the observed weak scale, which is critical not to the crossover at $v_h=0$ but to the sharp transition at ${\sim}\Lambda_{\rm QCD}$.
\end{abstract}

\maketitle

%%%%%%%%%%%%%%%%%%%%%%

\section{Introduction}
 
The Planck weak-scale hierarchy may be addressed by the near criticality of the Higgs mass parameter~\cite{Giudice:2006sn,Giudice:2008bi}. 
In this viewpoint, the small weak scale close to zero is special because the universe transitions between broken and unbroken phases of the electroweak symmetry at zero. The transition could generate various standard model (SM) backreactions that allow  dynamical selection of the weak scale~\cite{Graham:2015cka,Espinosa:2015eda,You:2017kah,Geller:2018xvz,Cheung:2018xnu,Arkani-Hamed:2020yna}. However, this transition is a second-order crossover in the SM, providing only relatively smooth selection rules.
In addition, the SM Higgs potential, once renormalization-group evolved, was found to yield another almost degenerate vacuum near the Planck scale~\cite{Degrassi:2012ry,Buttazzo:2013uya}. This surprising coincidence provides more evidence that the particular (seemingly unnatural) value of the weak scale might be related to criticality. This motivated ideas of multiple-point principle~\cite{Froggatt:1995rt,Froggatt:2001pa,Kawai:2013wwa,Hamada:2015dja,Kawai:2021lam}, classical scale invariance~\cite{Chun:2013soa,Chway:2013fzr,Hashimoto:2013hta}, as well as Higgs  inflation~\cite{Bezrukov:2007ep,Hamada:2013mya,Hamada:2014wna}. Extremely small dark energy is also thought to be near the critical point. But yet, whether and how criticality plays a crucial role in naturalness remain unclear.

Recently, a cosmological selection mechanism for criticality was developed in Ref.~\cite{Giudice:2021viw}, where inflationary quantum-dominated evolution of the relaxion inevitably drives a theory parameter close to a quantum critical point. In one setup, it is crucial that the critical point be the first-order separation between discrete phases with a significant energy difference, so that the Hubble rate can be sharply largest there. Then after long enough inflation (essentially eternal as will be discussed), such Hubble patches having a theory near the critical point will dominate the multiverse, as they are expanding and are reproduced most rapidly -- Hubble selection of the universe. This mechanism realizes self-organized criticality~\cite{Giudice:2021viw}; see also Refs.~\cite{Kartvelishvili:2020thd,Khoury:2019yoo}.

Then we ask the following: Why is the selection of criticality the selection of the observed small weak scale? To what first-order critical points does the Higgs mass  have relevance? 
Reference \cite{Giudice:2021viw} analyzed
the aforementioned renormalization-scale dependence of the SM Higgs vacuum structure~\cite{Degrassi:2012ry,Buttazzo:2013uya}, but the critical scale was found to be far above the weak scale.
References \cite{Geller:2018xvz,You:2017kah} studied a prototype model with multiple axions, where a QCD barrier trapping an axion disappears when $v_h$ turns off, so that the axion suddenly rolls down to the minimum, generating a large energy contrast necessary for Hubble selection. Critical changes of a theory could also induce the small weak scale in association with much smaller dark energy~\cite{Cheung:2018xnu,Arkani-Hamed:2020yna}.

In this Letter, we present a cosmological account of the weak scale from possible first-order zero-temperature (hence, quantum) critical points of QCD.\footnote{The small weak scale, if not due to criticalities or symmetries, could also be a result of the cosmological selection of anthropic~\cite{Cheung:2018xnu,Giudice:2019iwl,Strumia:2020bdy,Csaki:2020zqz,TitoDAgnolo:2021nhd,TitoDAgnolo:2021pjo} or entropic~\cite{Dvali:2003br,Dvali:2004tma,Kawai:2011qb,Hamada:2014xra,Arkani-Hamed:2016rle,Arvanitaki:2016xds} principles in a multiverse.} 
We first point out that QCD may have built-in quantum critical points at some $v_h^* \lesssim v_{\rm EW} = 246 \,{\rm GeV}$; this has yet to be studied, and we initiate an exploration using the three-flavor linear sigma model (LSM) of low-energy QCD. Then we present a relaxion model that realizes Hubble selection of the QCD criticality and self-organizes  $v_h$ close to the observed value. Then, $v_h$ is critical to ${\sim} \Lambda_{\rm QCD}$ (not to the crossover at zero). An added benefit is that the weak scale and $\Lambda_{\rm QCD}$ are generically close, which otherwise is accidental. 
Furthermore, building upon earlier works, we elaborate Hubble selection with different semi quantitative derivations.

We are inspired by observations that if the strange quark were slightly lighter, the (finite-temperature $T$) QCD chiral phase transition could have been first order. Although not yet firmly established~\cite{Brown:1990ev,Gavin:1993yk,DeTar:2009ef,Resch:2017vjs,deForcrand:2017cgb,Li:2017aki,Cuteri:2017zcb,Kuramashi:2020meg}, this possibility has been expected based on the (non)existence of infrared fixed points in the three-dimensional LSM~\cite{Pisarski:1983ms,Wilczek:1992sf}. 
In other words, QCD at $T=0$ too (relevant during inflation) may have a rich vacuum structure, as a  function of variable quark masses or $v_h$. Our initial phenomenological exploration of the vacuum structure shall be verified by dedicated research.

The Letter discusses the basic model ingredients (\Sec{sec:model}), exploration of QCD quantum critical points (\Sec{sec:QCD}), Hubble selection (\Sec{sec:HubbleSelection}), realization of the weak-scale criticality (\Sec{sec:weak}), and conclusions with future improvements.

%%%%%%%%
\section{Model}  \label{sec:model}

The model consists of the relaxion $\phi$~\cite{Graham:2015cka}, the Higgs $h$, and the meson field $\Sigma$: $V_{\rm tot} \= V_\phi + V_h + V_\Sigma$. The relaxion couples only to the Higgs sector, scanning $v_h$. But the change of $v_h$ induces changes in the $\Sigma$ sector, developing the desired quantum criticality at $v_h^*$.
Then the Hubble selection (acting on $\phi$) self-organizes the universe to the critical point.

The real-scalar relaxion potential is axion like:
\beq
V_\phi \= \Lambda_\phi^4 \cos \frac{\phi}{f_\phi}.
\eeq
For Hubble selection, its field range $f_\phi$ shall exceed the Planck scale (see later), which is possible with multiple axions~\cite{Kim:2004rp,Choi:2015fiu,Kaplan:2015fuy}.

The Higgs potential takes the SM form ($\lambda_h \simeq 0.13$) plus the coupling to the relaxion (see Ref. \cite{Graham:2015cka} for details)
\beq
V_h \= \frac{1}{2}(M^2 - g \widetilde{\phi}) h^2 + \frac{\lambda_h}{4} h^4 \, \, \to \, \, - \frac{1}{2}(g \phi) h^2 + \frac{\lambda_h}{4} h^4,
\eeq
where $h$ is the real Higgs field in unitary gauge. We shift $\phi$ such that the quadratic term $\mu_h^2 = -g\phi$ vanishes at $\phi {=} 0$. $v_h^2 \equiv -\mu_h^2/\lambda_h = g\phi/\lambda_h$ is used to label the relaxion scanning ($\vEW \,{=}\,246$ GeV).\footnote{QCD backreaction $V_h \ni y_q h \langle \bar{q} q \rangle /\sqrt{2}$ is ignored, as it is sizable only for the $SU(3)_V$ vacuum which is not Hubble selected for $v_h \lesssim v_h^*$.}
The required field range of $\phi$ to scan $\mu_h^2$ up to the cutoff $M^2$ is $\delta \phi \sim M^2/g$, thus we set 
$f_\phi = M^2/g.$
The dimensionful coupling $g$ is a spurion of the relaxion shift symmetry, and thus can be small naturally.

Below $\Lambda_{\rm QCD}=200 \MeV$, meson fields $\Sigma_{ij}(x)$
are relevant degrees of freedom, whose condensates are order parameters for chiral symmetry breaking. This vacuum structure as a function of $v_h$ is what we want to explore. It can be conveniently described by the LSM with $U(N_f)_L \times U(N_f)_R$ symmetry of QCD~\cite{Levy:1967,Gell-Mann:1960mvl,Lee:1970},
\bea
V_\Sigma &\,=\,& \mu^2 \Tr[ \Sigma \Sigma^\dagger ] + \lambda_1 ( \Tr[ \Sigma \Sigma^\dagger ])^2 + \lambda_2 \Tr[ (\Sigma \Sigma^\dagger)^2] \nonumber\\
&& - c( \det \Sigma + \det \Sigma^\dagger ) - \Tr [{\cal H} (\Sigma + \Sigma^\dagger)],
\label{eq:L} \eea
where fields and parameters are decomposed as
$\Sigma = (\sigma_a + i \pi_a) T^a, \, {\cal H} = h_a T^a$
with generators $T^a$ satisfying $\Tr [T^a T^b] = \delta^{ab}/2$ for $a=0, ..., N_f^2-1$. Without losing generality, $\lambda_{1,2}, h_a$ are real, $c>0$, and $\mu^2$ can take either sign.
$\Sigma$ is bifundamental under the symmetry. The first line of \Eq{eq:L} conserves $SU(N_f)_L \times SU(N_f)_R$; $\lambda_2$ is nonzero, otherwise symmetry is enhanced to $O(2N_f^2)$. One of the remaining $U(1)$'s is identified as the conserved baryon number $U(1)_V$, simply omitted in our discussion. The other $U(1)_A$ is anomalous, broken by the instanton contribution $c$ down to $Z_A(N_f)$~\cite{tHooft:1976rip,tHooft:1976snw}. Symmetries are further broken by ${\cal H}$, the leading chiral-symmetry-breaking mass term. We fix $N_f=3$ with the isospin symmetry $m_u = m_d$, as a first exploration; only $h_0,h_8 \ne 0$.

It is worthwhile to note that the LSM indeed possesses necessary features for quantum critical points. For $N_f=3$, the instanton term is a cubic potential, possibly creating local vacua (even with $\mu^2>0$).
The linear term ${\cal H}$ can destabilize the local vacua at critical quark masses or $v_h^*$, just as the external magnetic field (the linear term) in ferromagnets flips higher-energy spin directions at a critical field strength.

%%%
\begin{table*}[t] \centering
\begin{tabular}{ c || c  c   c c c c c | c c c}
%\hline \hline
parameter  & $f_\pi$ & $f_K$ & $m_\pi$ & $m_K$ & $m_\eta$ & $m_{\eta^\prime}$ & $m_{a_0}$   &$m_{f_0(500)}$ & $m_{f_0(1370)}$ & $m_{K^*(1430)}$ \\
\hline
measured  & 92.4 & 113  &  $139.57^{\pm 0.005}$ & $497.61^{ \pm 0.013}$ & $547.86^{ \pm 0.017}$ & $957.78^{ \pm 0.06}$ & $980^{ \pm 20}$    &$500^{ \pm 150}$ & $1370^{ \pm 150}$ & $1425^{ \pm 50}$  \\
benchmark & 92.4 & 113 & 137 & 491 & 534 & 973 & 1050    &731 & 1260 & 1140 \\
%\hline \hline
\end{tabular}
\caption{Predictions of the benchmark SM point [\Eq{eq:hSM} and (\ref{eq:SMpoint})], compared with data from the Particle Data Group~\cite{Zyla:2020zbs}. 
In units of $\MeV$.}
\label{tab:data} \end{table*}
%%%

%%%%%%%%%%%%%
\section{QCD quantum critical points}  \label{sec:QCD}

To explore the vacuum structure as a function of $v_h$, we first fix the benchmark ``SM point'' parameters of $V_\Sigma$, reproducing a measured meson spectrum, and then we deduce how these parameters change with $v_h$.

The masses of pions and kaons, being pseudo-Goldstones, are given by symmetry-breaking terms ${\cal H}$, related by partially conserved axial-vector currents,
\beq
\partial_\mu j^{\mu 5}_a \= m_{\pi_a}^2 f_{\pi_a} \pi_a \= \pi_b h_c d_{abc},
\label{eq:pcac0} \eeq
where the last equality is  obtained by the variation of  $V_\Sigma$ under chiral transformations. 
For pions $\pi^0 = \pi_3$ with $d_{3b0} = \sqrt{2/3}\delta_{b3}, d_{3b8} = 1/\sqrt{3} \delta_{b3}$, and for kaons $K^0 = (\pi_6 +i \pi_7)/\sqrt{2}$ with $d_{Kb0} = \sqrt{2/3} \delta_{bK}, d_{Kb8} = -1/\sqrt{12} \delta_{bK}$, we have
\beq
m_\pi^2 f_\pi =  \sqrt{\frac{2}{3}}h_0  + \frac{h_8}{\sqrt{3}}, \quad m_K^2 f_K =  \sqrt{\frac{2}{3}}h_0  - \frac{h_8}{2\sqrt{3}}.
\label{eq:PCAC} \eeq
Using  measured values of $m_{\pi,K}$ (Table~\ref{tab:data}), we fix the SM-point value of $h_{0,8}$~\cite{Fejos:2016hbp},
\beq
h_0(v_{\rm EW}) = (287 \MeV)^3, \,\,\,\, h_8(v_{\rm EW}) = -(312 \MeV)^3.
\label{eq:hSM} \eeq

We proceed to fit masses of other pseudoscalar and scalar mesons to data. The minima of $V_\Sigma$ are numerically found by considering the stability along all 18 field directions.
The $N_f=3$ LSM is known to have three types of vacua at ${\cal H}=0$~\cite{Lenaghan:2000ey,Bai:2017zhj}: $SU(3)_L \times SU(3)_R$ ($s_1 = s_3=0$), $SU(3)_V$ ($s_1=s_3 \ne0$), and $SU(2)_L \times SU(2)_R \times U(1)_V$ ($s_1=0, s_3\ne0$), where 
$\langle \Sigma \rangle \= \sigma_0 T^0 + \sigma_8 T^8 \= {\rm diag}( s_1, s_1, s_3)$.
In particular, the global $SU(3)_V$ vacuum (that we live today) and the local $SU(3)_L \times SU(3)_R$ vacuum \emph{coexist} if $\mu^2>0$ and
$K \, \equiv \, \frac{c^2}{2 \mu^2(3\lambda_1 + \lambda_2)} \, > \, 4.5$ with $3\lambda_1 + \lambda_2 >0$~\cite{Bai:2017zhj}.
Thus, this parameter space is our focus, that potentially exhibits first-order quantum critical points.

By scanning with these constraints, we found a range of good parameter space (see Appendix A in Supplemental Material \cite{Supp}).
The benchmark SM point is [with \Eq{eq:hSM}]
\beq
\mu^2 = (60 \,{\rm MeV})^2, \,\,\, c = 4800 \MeV, \,\,\, \lambda_1 = 7, \,\,\, \lambda_2 = 46,
\label{eq:SMpoint} \eeq
yielding $K=47.8$. Its goodness of fit to the meson spectrum (Table~\ref{tab:data}) is $\chi^2$/degrees of freedom = 0.44 with the first seven observables and 3.11 including all.
The first seven are the most reliable, while the last three are less precisely measured with unclear identities~\cite{Zyla:2020zbs}. Here, 5\% theoretical uncertainties are added as typical sizes of the perturbative corrections. Our benchmark is as good as existing benchmarks in the literature: Ref.~\cite{Lenaghan:2000ey} (with $\mu^2>0$) yielded 0.20 and 2.84, respectively, and Ref.~\cite{MeyerOrtmanns:1994nt} ($\mu^2<0$) yielded 1.22 and 4.64.\footnote{Large uncertainties of the last three observables allow various good fits; otherwise, the LSM would have been over constrained at the tree level. Higher-order, nonperturbative, and other effects may be calculated by lattice simulations~\cite{Gross:1980br,Pawlowski:1996ch,Heller:2015box,Braun:2020mhk,Dupuis:2020fhh}.}.
But ours differs in that quantum critical points $v_h^*$ may exist.

%%%%%%%%

We turn to discuss the vacuum structure away from the SM point for $v_h < \vEW$. How does $V_\Sigma$, in particular ${\cal H}$, depend on $v_h$? Since this is not known, we deduce it as follows. The current divergence [\Eq{eq:pcac0}] calculated from QCD or chiral Lagrangian yields $m_\pi^2 \propto m_q$, which is also $\propto h_{0,8}$ from \Eq{eq:PCAC}. It suggests $h_{0,8} \propto v_h$. Indeed, identifying the ${\cal H}$ term and the current mass term, ${\cal L} \ni -m_q (\bar{u} u + \bar{d} d) - m_s \bar{s}s$, yields the ratio 
$h_8/h_0 \simeq \left(  \frac{m_q - m_s}{\sqrt{3}}  \right) / \left(  \frac{2m_q + m_s}{\sqrt{6}}  \right) \simeq -1.3$ using $m_q = \frac{(m_u + m_d)}{2} = 3.4 \MeV$ and $m_s = 93 \MeV$~\cite{Zyla:2020zbs}, 
same as the ratio from \Eq{eq:hSM}.
Thus, we assume that ${\cal H}$ is linear to $v_h$ as
\beq
h_{0,8}(v_h) \= h_{0,8} (v_{\rm EW}) \frac{v_h}{v_{\rm EW}}.
\label{eq:Hscan} \eeq
Other LSM parameters and dimensionful factors could also depend on $v_h$, either directly or indirectly, e.g. via condensation or $\Lambda_{\rm QCD}$.
$\Lambda_{\rm QCD}$ depends on quark masses via renormalization running but only logarithmically, and instanton contributions on masses and condensates but complicated and nonperturbative~\cite{Schafer:1996wv}. 
In this initial exploration, \Eq{eq:Hscan} is assumed to be the only change of $V_\Sigma$ induced by the scanning of $v_h$.

\begin{figure}[t] \centering
\includegraphics[width=0.84\linewidth]{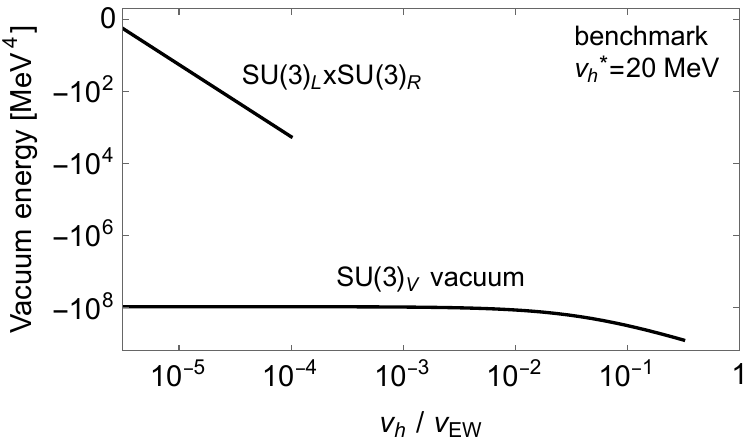}
\caption{Vacuum energies of benchmark coexisting QCD vacua at $T=0$,  as functions of $v_h$. The critical point is at $v_h^* \simeq 20 \MeV$, with a $\LQCD$-scale energy difference.
\label{fig:QCDvac} }
\end{figure}

The vacuum structure as a function of $v_h$ (other parameters fixed to the benchmark) is shown in \Fig{fig:QCDvac}. As $K>4.5$ dictates, there are coexisting vacua at $v_h=0$.
As $v_h$ (hence, ${\cal H}$) increases with \Eq{eq:Hscan}, and the metastable $SU(3)_L \times SU(3)_R$ vacuum becomes shallower and unstable at the critical point, which is found at $v_h^* \simeq 20 \MeV$; in fact, a wider range of $v_h^* = {\cal O}(1\sim100) \MeV$ is consistent with the meson data (see Appendix A in Supplemental Material \cite{Supp}). 
The energy difference of the coexisting vacua at the critical point is $93 \MeV$, comparable to $\LQCD$;
the potential energies are parameterized in \Eq{eq:LSMvacE1} and (\ref{eq:LSMvacE2}). 

In all, we have shown that QCD may possess quantum critical points at $v_h^* < v_{\rm EW}$, which needs dedicated verification.

%%

%%%%%%%%

\section{Hubble selection} \label{sec:HubbleSelection}

Inflationary quantum fluctuations on the scalar field allow access to a higher potential regime, which is forbidden classically. Although the field in each Hubble patch always rolls down in average, larger Hubble rates at higher potentials can make a difference in the global field-value distribution among patches, culminating in Hubble selection.
This section reviews and supplements it~\cite{Giudice:2021viw}.

The volume-weighted (global) distribution $\rho(\phi,t)$ of the field value $\phi$ obeys the modified Fokker-Planck equation (FPV)~\cite{Nakao:1988yi, Sasaki:1988df, Mijic:1990qx, Linde:1993nz}
\beq
\frac{\partial \rho(\phi,t)}{\partial t} \= \frac{\partial}{\partial \phi} \left( \frac{V^\prime}{3H} \rho \right) \+ \frac{1}{8\pi^2} \frac{\partial^2 (H^3 \rho) }{\partial \phi^2} \+ 3 \Delta H \rho.
\label{eq:FPV} \eeq
The first two terms represent the flow and diffusion, just as in the original Fokker-Planck equation which averages over Langevin motions. The variation of the Hubble rate $3\Delta H(\phi) =  \frac{V(\phi)}{2 \Mpl^2 H_0} \ll 3H_0$ accounts for volume weights within a distribution: $\Mpl = 2.4 \times 10^{18} \GeV$.
The meanings become clearer if we look at a solution (for a linear potential without boundary conditions),
\beq
\rho(\phi,t) \propto 
\exp \left\{ \frac{-1}{2\sigma_\phi^2(t)} \Big[ \phi - \big( \phi_0 + \dot{\phi}_c t + \tfrac{3 }{2}(\Delta H)^\prime \sigma_\phi^2 t \big) \Big]^2 \right\},
\label{eq:FPVsol0} \eeq
where the exponent describes the motion of a peak. $\dot{\phi}_c = -V^\prime/3H$ is classical rolling. Remarkably, an additional velocity $\dot{\phi}_H=3(\Delta H)^\prime \sigma_\phi(t)^2$ with opposite sign arises from volume weights within the width $\sigma_\phi$, which grows in the beginning of FPV evolutions due to quantum diffusion $\sigma_\phi^2(t) = (\frac{H}{2\pi})^2 Ht$ from the de Sitter temperature $H/2\pi$~\cite{Starobinsky:1994bd, Gibbons:1977mu}. 

``Hubble selection'' starts to operate when the peak of a distribution starts to \emph{climb up} the potential:
$\sigma_\phi^2 \simeq \frac{2}{3}\Mpl^2$.
The width at this moment is always Planckian, reflecting its quantum nature.
The field excursion by this moment is non-negligible,
$\Delta \phi \sim \frac{4 \pi^2 \Mpl}{9} \frac{\Lambda_\phi^4}{H^4} \frac{ \Mpl}{M^2/g}$.
Thus, for a peak to climb, the field range $\delta \phi \sim M^2/g$ (needed to scan $\mu_h^2$ up to $M^2$) must accommodate both the field excursion $\Delta \phi$ (stronger condition) and width $\sigma_\phi$, yielding respectively
\beq
g \,\lesssim \,  H \frac{H}{\Mpl} \frac{M^2}{\Lambda_\phi^2} \,  \lesssim \, \frac{M^2}{\Mpl}.
\label{eq:globalQBC}\eeq
We call this condition \emph{global} quantum beats classical (QBC). It is stronger than the usual local QBC,
$V^\prime \lesssim H^3$, requiring $g \lesssim H \frac{H^2}{\Lambda_\phi^2}\frac{M^2}{\Lambda_\phi^2}$,
because $\Lambda_\phi^2 \lesssim H \Mpl$ from condition 1 later. It also has different meanings as it involves the field range while the local one depends only on the potential slope. It turns out to be equivalent to the Quantum+Volume (QV) condition in Ref.~\cite{Giudice:2021viw} (see Appendix B in Supplemental Material \cite{Supp}) which also accounted for volume effects. 
If it is not satisfied, $\rho$ makes an equilibrium at the bottom of a potential, but with the sub-Planckian width consistently $\sigma_\phi \sim H^2/m_\phi \sim H^2M^2/\Lambda_\phi^2 g \lesssim \Mpl$~\cite{Graham:2018jyp,Takahashi:2018tdu}.
Thus, we require the global QBC \Eq{eq:globalQBC} for Hubble selection.

The e-folding until this moment
$\Delta N \simeq \frac{8 \pi^2 \Mpl^2}{3 H^2}$
already saturates the upper bound for finite inflation, 
$\frac{2\pi^2}{3}\frac{\Mpl^2}{H^2}$,
given by the de Sitter entropy~\cite{ArkaniHamed:2007ky,Dubovsky:2008rf,Dubovsky:2011uy}.
Thus, Hubble selection needs eternal inflation, and the universe eventually reaches a stationary state~\cite{Aryal:1987vn,Nambu:1989uf,Linde:1993xx,Garcia-Bellido:1994gng}. 
Probability distributions are to be defined within an ensemble of Hubble patches that have reached reheating~\cite{Vilenkin:1995yd,Vilenkin:1998kr,Creminelli:2008es}. As the latest patches dominate the ensemble with an exponentially larger number, only stationary or equilibrium distributions matter; for landscapes, this can be different~\cite{Freivogel:2011eg,Denef:2017cxt,Khoury:2019ajl,Khoury:2021grg}.

$\rho(\phi,t)$ makes an equilibrium somewhere near the top of a potential, which is the critical point $\phi_*$ in this work. The distribution can be especially narrow [Planckian in the global QBC regime; see Eq.(B6) in Supplemental Material \cite{Supp}] if energy drops sharply after $\phi_*$.
This is how Hubble selection self-organizes the universe toward critical points~\cite{Giudice:2021viw}.

The flatter the potential is (with stronger quantum effects), the closer to $\phi_*$ is the equilibrium. The closest possible field distance is Planckian, again reflecting the uncertainty principle. 
For even flatter potentials, the equilibrium distribution rather spreads away from $\phi_*$, because distributions will be flat in the limit $V^\prime\to 0$. The equilibrium near $\phi_*$ is estimated as follows.
The boundary condition $\rho(\phi \geq \phi_*) =0$ (discarding Hubble patches with $\phi \geq \phi_*$) induces repulsive motion $\dot{\phi}_b \sim - H^3/(8\pi^2 \sigma_\phi)$~\cite{AppB}, so that the balance requires
$\dot{\phi}_c + \dot{\phi}_H + \dot{\phi}_b \simeq 0$.
When $\dot{\phi}_b \gg \dot{\phi}_c$, an equilibrium is reached with
\beq
\sigma_\phi \,\simeq\, \left( \frac{ \Mpl^2 H^4}{4\pi^2 V^\prime} \right)^{1/3} \,\simeq\, \left( \frac{ \Mpl^2 M^2 H^4}{4\pi^2 g \Lambda_\phi^4} \right)^{1/3},
\label{eq:Q2Vbalance}\eeq
which is the width in the Quantum${}^2$+Volume (\QtoV) regime~\cite{Giudice:2021viw}. 
The width indeed increases as $V$ flattens; nevertheless, the $v_h$ distribution can be arbitrarily narrowed, as will be discussed. One also expects $|\phi_{\rm peak}-\phi_*| \sim \sigma_\phi$ from dimensional ground.
These heuristic discussions on \QtoV are demonstrated with the method of images in Appendix B of Supplemental Material \cite{Supp}.

A theory enters the \QtoV regime when the balance width becomes larger than $\Mpl$ (the width in the global QBC):
\beq
g \,\lesssim\, H \frac{H}{\Mpl} \frac{H^2 M^2}{\Lambda_\phi^4}.
\label{eq:Q2V}\eeq
This is equivalent to $V^\prime \lesssim H^3 H/\Mpl$ \cite{Giudice:2021viw}, which is also derived from the local balance near $\phi_*$. 
\QtoV is typically stronger than the global QBC and not absolutely needed for Hubble selection, but later will be useful for efficient localization of $v_h$.

%%%%%%%%
\section{The weak scale criticality} \label{sec:weak}
 
Finally, we come  to calculate the equilibrium distribution of $\rho(v_h)$ in our model. 
We first discuss conditions for the successful Hubble selection of $v_h^*$, and then present benchmark results.

The scanning of $v_h$ starts by $\phi$ rolling up its potential from $\phi<0$ to $>0$.
When $\phi<0$ ($\mu_h^2 > 0$), $v_h=0$ and $V_h=V_\Sigma =0$ remain unchanged with $\phi$. Thus $\phi$ simply keeps growing, driven by quantum effects.
The only constraint is that $V_\phi$ must  not affect the inflation dynamics (condition 1):
$\Lambda_\phi^4  \, \lesssim \, H^2 \Mpl^2.$

As soon as $\phi>0$ ($\mu_h^2 \leq 0$), the Higgs gets the vacuum expectation value $v_h > 0$, and $V_h, V_\Sigma$ minima now evolve with $\phi$.
$V_h = -\frac{\lambda_h}{4} v_h^4$,
and coexisting vacua of $V_\Sigma$ are, at leading orders in $v_h$,
\bea
V_{\Sigma (L \times R)} &\, \simeq \,& - a_1 \LQCD^4 \frac{v_h^2 }{ \vEW^2 }, \label{eq:LSMvacE1} \\
V_{\Sigma(V)} &\, \simeq \,& - \LQCD^4( a_2 + a_3 \frac{v_h}{\vEW}),
\label{eq:LSMvacE2} \eea
where $a_1 \simeq 114, a_2 \simeq 0.059, a_3 \simeq 1.38$ for the benchmark (\Fig{fig:QCDvac}). $\langle \sigma_0 \rangle_{(L \times R)} \simeq - \langle \sigma_8 \rangle \simeq a_4 v_h$ with $a_4\simeq 0.025$.
Note that $V_{h,\Sigma}$ decrease with $\phi$, which must be slower than the increase of $V_\phi  \, \simeq \, \frac{ g \Lambda_\phi^4}{M^2} \phi$, for Hubble selection.

\begin{figure}[t] \centering
\includegraphics[width=0.84\linewidth]{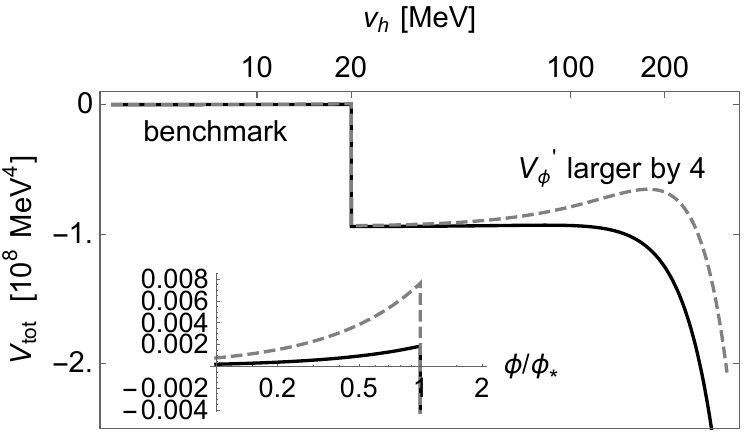}
\includegraphics[width=0.84\linewidth]{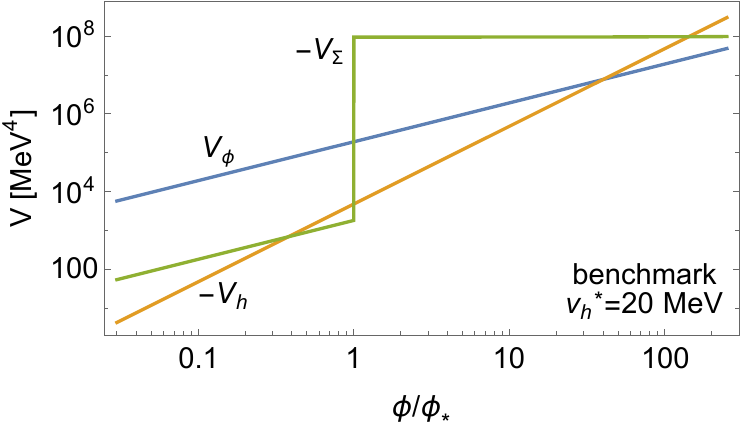}
\caption{Top: Total potential energy near the critical point as a function of $\phi$, for the benchmark \Eq{eq:benchmark}; dashed line for comparison. Inset: Zoom-in near $\phi_*$.
Bottom: Individual contribution from $V_\phi$, $|V_\Sigma|$, and $|V_h|$. 
\label{fig:pot} }
\end{figure}

Which potential dominates the $\phi$ dynamics? \Fig{fig:pot} shows individual potential with $\phi$, whose slope is
\bea
\frac{\delta V_\phi}{\delta \phi} &\,\sim\,& g\frac{\Lambda_\phi^4}{M^2},  \quad  \frac{ \delta V_{\Sigma(L\times R)}}{\delta \phi} \,\sim\, -a_1 \frac{g}{\lambda_h} \frac{\LQCD^4}{\vEW^2},   \label{eq:Vwrtphi}\\ 
\frac{\delta V_h}{\delta \phi} &\,\sim\,& - \frac{g}{2} v_h^2,  \quad  \frac{ \delta V_{\Sigma(V)}}{\delta \phi} \,\sim\, -a_3 \frac{g}{\lambda_h} \frac{\LQCD^4}{v_h \vEW}. \nonumber
\eea
The dominance of growing $\delta V_\phi/\delta \phi$ up to $v_h \leq v_h^*$ requires
$\Lambda_\phi^2/ M \gtrsim v_h^*$, unless $v_h$ is too small. After $\Lambda_{\rm QCD}$-scale energy drops in $V_\Sigma$ at $v_h^*$, dominant $V_\phi$ keeps growing. For large enough $v_h \gtrsim \Lambda_{\rm QCD} (\gtrsim v_h^*)$, decreasing $V_h$ begins to dominate and is prohibited from being Hubble selected again. So we need to make sure that $V_\phi$ never compensates the energy drop in the intermediate region $v_h^* \lesssim v_h \lesssim \Lambda_{\rm QCD}$:
$\Delta V_\phi \,\simeq\, \frac{g \Lambda_\phi^4}{M^2} \frac{\lambda_h \LQCD^2}{g} \, \lesssim \, \LQCD^4.$
In all, $V_\phi$ cannot be too flat or too steep (condition 2):
\beq
v_h^* \lesssim  \Lambda_\phi^2 / M \lesssim \LQCD.
\label{eq:condM} \eeq

In addition, $h$ and $\Sigma$ are required to sit in their respective minima, not quantum driven to overflow their potentials. 
Their equilibrium widths must be small enough:
$H^2/m_{h,\Sigma} \lesssim \LQCD$, with $m_h \sim v_h$ and $m_\Sigma \sim \LQCD$
in the higher-energy $SU(3)_L \times SU(3)_R$ vacuum. 
Since $v_h^* \lesssim \LQCD$ from condition 2, we obtain (condition 3):
$H \,\lesssim \, v_h^*.$

For numerical studies, we use the following benchmark ($v_h^*$ from \Sec{sec:QCD}),
\bea
v_h^* &\,\simeq\,& 20 \MeV, \quad H \= v_h^*, \quad M \= 3\times 10^{-3} \Mpl, \nonumber\\
\Lambda_\phi^2 &\=& 10^{-2} H \Mpl, \quad g \= 10^{-3}H^2/\Mpl,
\label{eq:benchmark} \eea
satisfying the global QBC [\Eq{eq:globalQBC}](and \QtoV [\Eq{eq:Q2V}] marginally) and conditions 1--3. 
Potential energies near $v_h^*$ are shown in \Fig{fig:pot}.  As desired, the total energy peaks sharply at $v_h^*$, drops significantly, and is never compensated afterwards; for much smaller or larger $V_\phi^\prime$, energy would not sharply peak.

\begin{figure}[t] \centering
\includegraphics[width=0.84\linewidth]{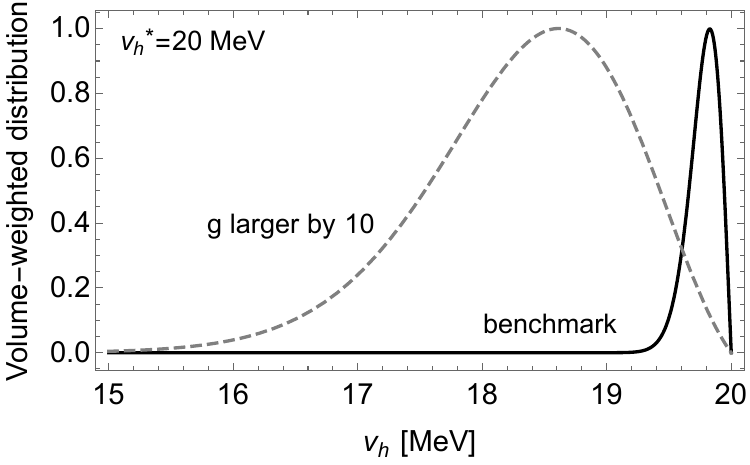}
\caption{The probability distribution of $v_h$ among Hubble patches that have reached reheating. $\sigma_{v_h} \simeq 0.1\MeV \ll v_h^*$ for the benchmark \Eq{eq:benchmark}; dashed line for comparison.
\label{fig:eqdist} }
\end{figure}

The large-time equilibrium distribution of $v_h$ is shown in \Fig{fig:eqdist}; see Appendix C in Supplemental Material \cite{Supp} for details.
The width $\sigma_{v_h}$ is translated from $\sigma_{\phi}$ via $v_h^2 = g\phi/\lambda_h$ as
\beq
\sigma_{v_h} \,\simeq\, \frac{g \sigma_\phi}{2 \lambda_h v_h^*},
\eeq
where $\sigma_\phi \simeq 1.3 \Mpl$ from \Eq{eq:Q2Vbalance} for the benchmark with marginal \QtoV.\footnote{The Planckian width is a generic result of the global QBC, 
$\sigma_\phi \simeq \phi_* \left( \frac{3 \phi_*^2}{2 \Mpl^2} \right)^{-1/2} \sim \Mpl$; see Eq. (B6) in Supplemental Material \cite{Supp}.}
Thus we have
$\sigma_{v_h} \simeq 0.1 \MeV \ll v_h^*$, which is narrow enough so that most Hubble patches self-organize to have $v_h\approx v_h^*$ ($\sim \vEW$).
Note that it can be arbitrarily narrower at the price of arbitrarily smaller $g$ or larger $\phi$ range, moving into a deeper \QtoV regime; only the resulting hierarchy $f_\phi=M^2/g \gg \Mpl$ needs to be generated consistently in field theory~\cite{Kim:2004rp,Choi:2015fiu,Kaplan:2015fuy}. On the other limit, unwanted $\sigma_{v_h} \gtrsim v_h^*$ is resulted for 10--100 times larger $g$, where \QtoV is not even marginally satisfied.

Postinflationary dynamics is model dependent but such that $\phi$ slow rolls to the today's $\vEW$.
Today, $\phi$ could be still safely slow rolling or trapped by SM backreactions. Signals from  time-dependent $v_h$ or phase transitions could be produced.

%%%%%%%%

\section{Discussion}

In this Letter, we have discussed the self-organized criticality of the weak scale, by exploiting possible first-order quantum critical points of QCD. Although we saw some success, our exploration of critical points is much simplified and far from conclusive. We have used only LSM with $N_f=3$ at the tree level with a simplified dependence on $v_h \leq \vEW$ in \Eq{eq:Hscan}. 
They shall be verified and generalized by lattice calculations~\cite{Brown:1990ev,Gavin:1993yk,DeTar:2009ef,Resch:2017vjs,deForcrand:2017cgb,Li:2017aki,Cuteri:2017zcb,Kuramashi:2020meg}, incorporating higher-order and nonperturbative effects~\cite{Gross:1980br,Pawlowski:1996ch,Heller:2015box,Braun:2020mhk,Dupuis:2020fhh}, not only for the SM point but also away from it with $v_h \leq \vEW$. $N_f >3$ likely yields a richer vacuum structure but needs a dedicated calculation. Theoretical intuitions from confining gauge theories might also be useful. If such a critical point is indeed built in QCD, it would shed significant light on the role of near criticality of the SM.

The proposed scenario makes an advancement on the hierarchy problem, albeit not yet completely solving it.
It is not complete because Hubble selection requires a mild separation of scales $\Lambda_\phi \ll M$ from \Eq{eq:condM} (if $M \sim \Lambda_\phi$ strictly, $M \lesssim \LQCD$ is too low) but this is not quantum stable (Higgs loop diagrams with external relaxion legs yield $\Lambda_\phi \sim$ cutoff $M$~\cite{Espinosa:2015eda,Choi:2015fiu}). Thus, a little hierarchy remains; with the fine-tuning $\epsilon \equiv \Lambda_\phi / M <1$, the cutoff can be as high as $M \lesssim \LQCD/\epsilon^2$.
Another advancement is that choosing $\Lambda_\phi$ can be translated to a dynamical problem of choosing dimensionless parameters of the extended relaxion sector, such as in Ref. \cite{Espinosa:2015eda}. Further explorations will be enlightening.

The near criticality and naturalness of nature may be intimately connected by quantum cosmology, with necessary criticality perhaps built in just around the SM. Further theoretical and experimental studies are encouraged to unveil this connection.

%%%%%
\medskip
\begin{acknowledgments}
We thank Sang Hui Im, Hyung Do Kim, Choonkyu Lee, and Ke-Pan Xie for valuable conversations. We are supported by National Research Foundation of Korea under Grant No. NRF-2019R1C1C1010050, and S.J. also by a POSCO Science Fellowship. 
\end{acknowledgments}

%%%%%%%%

\clearpage

\title{Supplemental Material: \\ Hubble selection of the weak scale from QCD quantum critical point}

\maketitle

\appendix
%%%%%%%%
\section{Quantum critical points of LSM} \label{app:LSM}

We present our initial exploration of quantum critical points $v_h^*$ of the $N_f=3$ LSM at tree-level. In Sec.~III, we presented the benchmark SM point, Eqs.~(6) and (7), that best fits the meson spectrum in Table I. Here we discuss further details of our search and the resulting ranges of best-fit parameters and $v_h^*$. 

It turns out that $\lambda_1$ and $\mu^2$ are least constrained by meson spectrum, as the last three meson observables in Table I have large uncertainties; without such freedom, the tree-level LSM would have been over-constrained.
So we vary these two parameters while fixing all others to the benchmark values.
We should also focus on the parameter space with $K>4.5$ so that coexisting vacua are present at ${\cal H}=0$; this roughly requires $\mu^2 \lesssim $ 10 times the benchmark value.

\begin{figure}[h] \centering
\includegraphics[width=0.84\linewidth]{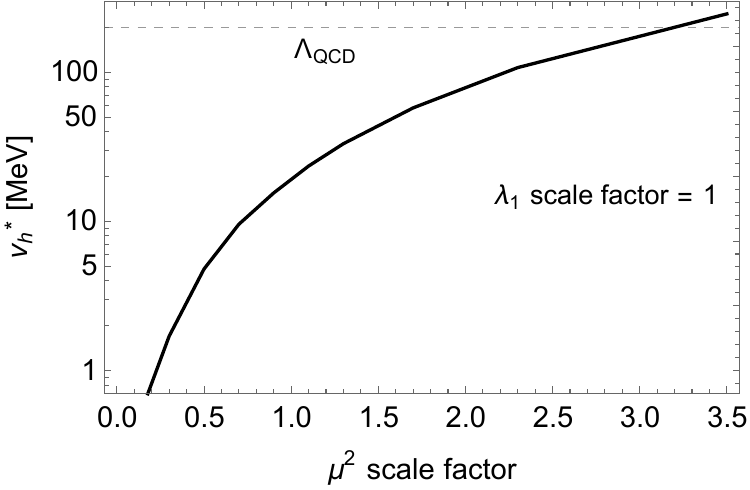}
\includegraphics[width=0.84\linewidth]{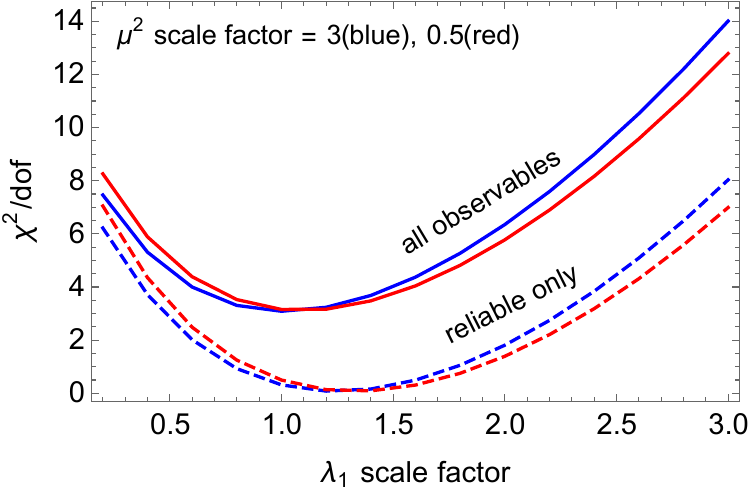}
\caption{(Top): Quantum critical point $v_h^*$ in the $N_f=3$ LSM. Other parameters are fixed to benchmark values; scale factors are relative to the benchmark values. (Bottom): $\chi^2$/dof of the global fit to meson spectrum in Table 1.
\label{fig:scan} }
\end{figure}

\Fig{fig:scan} shows the numerical results of the critical point $v_h^*$ (upper panel) and $\chi^2$/dof (lower) as a function of $\mu^2$ and $\lambda_1$; we denote the values of $\mu^2$ and $\lambda_1$ by the ratio (scale factors) relative to the benchmark values, as $\hat{\mu}^2$ and $\hat{\lambda}_1$. We found that $v_h^*$ is sensitive mostly only to $\mu^2$, while $\chi^2$ only to $\lambda_1$. 
Thus, a wide range of $v_h^*$ is found to be consistent. 

At the $2\sigma$ confidence level ($\chi^2$/dof = 4), $\hat{\lambda}_1= 0.6\sim1.5$ is consistent with all meson observables in Table I, and $\hat{\lambda}_1=0.4 \sim 2.5$ is consistent with first seven observables. Fixing $\hat{\lambda}_1=1$ (or $\lambda_1=7$) and restricting $\hat{\mu}^2 < 10$ (for $K>4.5$), any values $v_h^* = {\cal O}(1 \sim 100)$ MeV are allowed for some $\hat{\mu}^2$. Specifically, $v_h^* = \Lambda_{\rm QCD} $ for $\hat{\mu}^2 \simeq 3$, and $v_h^* = 1$ MeV for $\hat{\mu}^2 \simeq 0.3$. Although even smaller $v_h^*$ is allowed with smaller $\mu^2$, we do not want too small $v_h^*$ farther away from $\vEW$.
In all, as alluded, $v_h^* = {\cal O}(1-100) \MeV$ can be obtained in the parameter space consistent with meson spectrum, spanned by ${\cal O}(1)$ variations of $\mu^2$. 

The energy contrast at the critical point does not vary much, 70$\sim$120 MeV close to $\LQCD$, in the parameter space shown in the figure.

%

%%%%%%%%
\section{Quantum regimes} \label{app:quantreg}

In Sec.~IV, we showed that quantum-dominated evolution falls into two regimes: the global QBC and \QtoV. In this appendix, we discuss further details. We first introduce the QV and \QtoV regimes derived in \cite{Giudice:2021viw} and show their equivalence to our results; then we discuss the scaling behaviors of equilibrium solutions, which turn out to provide further intuition as well as useful handles in numerical calculations (see \App{app:eqdist}); finally we derive equilibrium properties of \QtoV near a boundary using the method of images. 
We also refer to \cite{Giudice:2021viw,Graham:2018jyp,Cheung:2018xnu} and references therein for other details on FPV solutions.

%%%%%
\subsection{Equivalence of quantum regimes}

Consider a general class of potential
\beq
V(\varphi) \=  V_0 \, \omega(\varphi), \qquad  |\varphi| \leq 1
\label{eq:genpot} \eeq
where the field $\varphi = \phi/\phi_0$ is normalized by the field range $\phi_0$. The potential height is $V_0 = V(\varphi=1)-V(\varphi=0)$ with $\omega(\varphi=1)=1$, $\omega(\varphi=0)=0$, and $d \omega / d\varphi ={\cal O}(1)$. Any potential (subdominant to the inflaton's) can be written in this form, by shifting the field and adjusting the potential height by a small constant $\ll 3 \Mpl^2 H_0^2$.

Up to leading orders of $V_0 \ll 3\Mpl^2 H_0^2$, FPV can be written as~\cite{Giudice:2021viw} (assuming $H_0 \ll \Mpl$) 
\beq
\frac{\partial \rho(\phi,t)}{\partial T} \=  \frac{d}{d\varphi} \left( \rho \frac{d\omega}{d\varphi} \right)+ \frac{\alpha}{2} \frac{d^2 \rho}{d \varphi^2}  + \beta \rho \omega, 
\label{eq:FPVre} \eeq
where ($'$ denotes the $\phi$ derivative)
\beq
\alpha \,\equiv\, \frac{3}{4\pi^2} \frac{H_0^4}{ V^\prime \phi_0}, \qquad \beta \,\equiv\, \frac{3 \phi_0^2}{2 \Mpl^2}
\eeq
with the dimensionless time  $T \equiv \frac{V^\prime}{3 H_0 \phi_0} t$ normalized by the characteristic time for classical relaxation $\sim \phi_0/\dot{\phi} \simeq 3H_0 \phi_0/V^\prime$. In a rough sense, $\alpha$ comes from the second-derivative diffusion term in FPV, measuring quantum effects; $\beta$ from the volume term, also measuring the scale of the relevant field range in Planck units.
Note that \Eq{eq:FPVre} depends mostly only on the potential slope, but not on the absolute height as long as it is much smaller than the inflaton potential; this allows to write in the form \Eq{eq:genpot}.
Then \cite{Giudice:2021viw} argued that the conditions for the QV and \QtoV regimes are, respectively,
\beq
\alpha \beta \,\gtrsim\, 1, \qquad \alpha^2 \beta \,\gtrsim\, 1.
\eeq

Applied to our model with $V_0 = \Lambda_\phi^4$ and $\phi_0 = f_\phi = M^2/g$ being the full field range, the two conditions read 
\bea
g \lesssim H \frac{H}{\Mpl}\frac{M^2}{\Lambda_\phi^2}, \qquad  g \lesssim H \frac{H}{\Mpl} \frac{H^2M^2}{\Lambda_\phi^4}.
\eea
These agree with the global QBC in Eq.~(11) (not with the local QBC) and the \QtoV derived in Eq.~(13). 
\cite{Giudice:2021viw} derived these conditions by requiring the positivity of a dominant eigenmode with absorbing (vanishing) boundary conditions. We rather ended up with the same conditions in yet other ways: the peak climbing up a linear potential, the balance between quantum climbing and boundary repulsion, and the balance width growing larger than Planckian (the minimum allowed by uncertainty principle). It also provided more dynamical explanation of why and how the field range is involved. 

Furthermore, since quantum dynamics is set by $\alpha$ and $\beta$, so are equilibrium properties. It was derived in~\cite{Giudice:2021viw} that the width of a localized equilibrium distribution is
\beq
\sigma_\phi \,\sim \, \phi_0 \beta^{-1/2} \,\simeq\, \Mpl, \qquad \sigma_\phi \,\sim\, \phi_0 \left( \frac{\alpha}{\beta} \right)^{1/3},
\label{eq:width-app}
\eeq
for the QV and \QtoV, respectively. The former always yields the Planckian width (as in our benchmark in Sec.~V), the minimum allowed by uncertainty principle; and the latter agrees with our Eq.~(12), derived from the local balance near $\phi_*$.

%%%%%
\subsection{Scaling of equilibrium solution} \label{app:scaling}

The key idea in this subsection is that if the equilibrium distribution is well localized, we should be able to get the same solution whether by considering a full field range in FPV or just a large enough range around the distribution. Since the distribution will be localized near the upper boundary (the critical point in this work), solutions must also be largely independent on the lower boundary conditions. Thus, properties of solutions, expressed in terms of $\alpha$ and $\beta$, must scale properly under the re-scaling of the field range considered in FPV.

When it comes to calculate the equilibrium distribution,
it is good enough to consider a linear potential, since $\rho(\phi)$ is localized.
Consider a linear potential (in \Eq{eq:genpot})
\beq
\omega(\varphi) \= \varphi.
\label{eq:linearpot-app} \eeq
$\varphi = 1$ is the upper boundary or the critical point. The field range $\phi_0$ does not have to be the full field range, but only needs to contain a large enough range around the critical point. 
$\alpha$ and $\beta$ for this case are
\beq
\alpha \= \frac{3}{4\pi^2} \frac{H^4}{V_0}, \qquad \beta \= \frac{3}{2} \frac{\phi_0^2}{\Mpl^2}.
\label{eq:alpha-app} \eeq

Under an arbitrary scaling of the field range by $a$ (with the upper boundary fixed at the critical point), $\phi_0 \to a \phi_0$ and $V_0 \to a V_0$, hence
\beq
\alpha  \,\propto\, a^{-1}, \qquad \beta \,\propto \, a^2.
\eeq
Thus, the QV and global QBC conditions, $\alpha \beta \propto a $, scale proportionally, reflecting again that they encode the necessary field range for Hubble selection. On the other hand, the \QtoV condition  is invariant $\alpha^2 \beta \propto a^0$, as it measures only the local balance near the upper boundary, as derived in Sec.~IV and \App{app:boundary}.

However, physical properties of equilibrium distributions must remain invariant under such arbitrary scaling. Indeed, the widths in \Eq{eq:width-app} are scale invariant as
\beq
\sigma_\phi \,\sim\, \phi_0 \beta^{-1/2} \,\propto\, a^0, \qquad \sigma_\phi \,\sim\, \phi_0 (\alpha/\beta)^{1/3} \,\propto\, a^0.
\eeq
The peak locations (derived in~\cite{Giudice:2021viw} and our \App{app:boundary}) are also scale invariant as
\beq
\phi_0 - \phi_{\rm peak}  \,\sim\, \frac{\phi_0}{2\alpha \beta} \,\simeq\, \frac{4\pi^2}{9} \frac{V^\prime \Mpl^2}{H^4} \,\propto \, a^0, \quad  \sim\, \sigma_\phi  \,\propto\, a^0,
\eeq
for the QV and \QtoV, respectively. Notably, the peak location in the QV regime is the same as the minimum field excursion $\Delta \phi$ needed for a peak to start climbing; this led to the global QBC in Eq.~(11).

What do these mean practically?  As long as $\alpha \beta \gtrsim 1$ and $\sigma_\varphi < 1$,  the same correct solutions can be obtained with any convenient field range around the critical point. If expressed in terms of $\varphi$, solutions look broader or narrower just due to different field normalization $a\phi_0$. But in terms of physical field value $\phi$, they are the same. This can be useful in numerically solving FPV; see \App{app:eqdist}.

\begin{figure}[t] \centering
\includegraphics[width=0.84\linewidth]{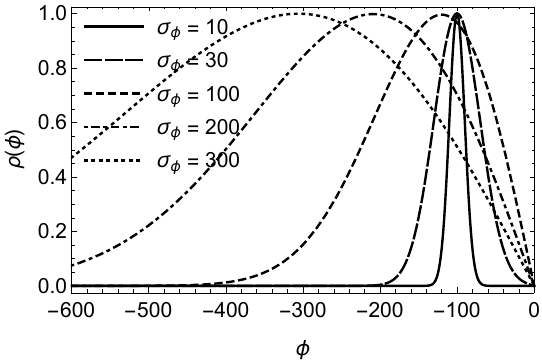}
\caption{Effects of the boundary at $\phi_0=0$ on the volume-weighted distribution $\rho(\phi)$. For several choices of $\sigma_\phi$ with fixed $\Delta \phi =100$ in the solution \Eq{eq:image} obtained by the method of images. As the boundary begins to matter (\QtoV), the distribution becomes pushed away.
\label{fig:boundary} }
\end{figure}
%

%%%%%
\subsection{Boundary effects} \label{app:boundary}

Lastly, we demonstrate the boundary effects in the \QtoV regime discussed heuristically in Sec.~IV. 

In solving FPV, the vanishing boundary condition at $\phi_0$ is relevant to our work, as any Hubble patches with $\phi > \phi_0$ (critical point) are out of Hubble selection, hence discarded.
The vanishing boundary condition can be imposed by the method of images. Putting two charges at $\phi_0 \mp \Delta \phi$ and ignoring Hubble expansion for simplicity, we can express the solution in the form~\cite{Creminelli:2008es} 
\beq
\rho(\phi) \, \propto \, e^{-\frac{[\phi - (\phi_0 - \Delta \phi)]^2}{2\sigma_\phi^2}} - e^{-\frac{[\phi - (\phi_0 + \Delta \phi)]^2}{2\sigma_\phi^2}},
\label{eq:image} \eeq
where $\phi <\phi_0$ is the physical region.
It describes that, as $\Delta \phi$ and $\sigma_\phi$ become comparable, or equivalently as the significant portion of the would-be distribution passes the boundary, the boundary becomes relevant, and the final distribution is displaced away from it. \Fig{fig:boundary} demonstrates this behavior as a function of $\sigma_\phi$ for fixed $\Delta \phi$. This effectively produces repulsive motion.

But physically, why and how does a boundary affect a peak located away from it? The boundary condition generates the asymmetric absorbing probability,
skewing the distribution, which otherwise would be symmetrically quantum diffused. 
The peak location $\phi_{\rm peak}$ is at the extremum satisfying
\beq
e^{\frac{2(\phi_0 - \phi_{\rm peak}) \Delta \phi}{\sigma_\phi^2}} \= \frac{(\phi_0 - \phi_{\rm peak}) + \Delta \phi}{(\phi_0 - \phi_{\rm peak}) - \Delta \phi}.
\eeq 
When $\sigma_\phi \gtrsim \Delta \phi$, the boundary is most important, and
\beq
\phi_0 - \phi_{\rm peak} \,\simeq \, \sigma_\phi. \label{eq:peakQ2V}
\eeq
This must be true as $\sigma_\phi$ is the only dimensionful parameter in this limit.  This effectively means repulsive motion
\beq
\dot{\phi}_b \, \sim \, - \frac{d \sigma_\phi}{dt} \= -\frac{H^3}{8\pi^2 \sigma_\phi},
\eeq
justifying the formula for the \QtoV regime used in Sec.~IV.

%%%%%%%%
\section{Numerical calculation of the equilibrium distribution}  \label{app:eqdist}

In this appendix, we summarize our numerical calculation of the equilibrium distribution $\rho(\phi)$. We use the solution given in~\cite{Giudice:2021viw}, obtained for a linear potential with the vanishing upper boundary condition, as discussed in \App{app:scaling}, and here we discuss useful technical steps involved. See also  \cite{Graham:2018jyp,Cheung:2018xnu} for similar ways to solve FPV at large time.

For the given $\alpha$ and $\beta$ in \Eq{eq:alpha-app} from a linear potential \Eq{eq:linearpot-app}, the equilibrium solution, up to a normalization, is given by~\cite{Giudice:2021viw}
\begin{equation}
\rho(\varphi) \,\propto\, e^{-\frac{\varphi}{\alpha}} \left[\frac{\text{Ai} (x(\varphi)) }{\text{Ai} (x(-1)) } - \frac{\text{Bi} (x(\varphi)) }{\text{Bi} (x(-1)) } \right],  
\end{equation}
where Ai and Bi are the Airy functions with 
\begin{equation}
x(\varphi) \,\equiv\, a_1 + \left(\frac{2\beta}{\alpha}\right)^{1/3} (1-\varphi).
\end{equation}
$a_1\simeq -2.3381$ denotes the first zero of $\text{Ai}$.
Since the distribution is highly localized, numerical evaluation of $\rho(\phi)$ often suffers technical difficulties. Close to the upper boundary $\varphi=1$, the solution can be approximated by ignoring small Bi contributions as
\beq
\rho(\varphi) \,\propto\, e^{-\frac{\varphi}{\alpha}} \text{Ai}(x(\varphi)).
\label{eq:eqsol} \eeq

For our benchmark Eq.~(18) with the upper boundary at the critical point $\phi_0 = \phi_* =  (v_h^*)^2 \lambda_h/g $ and $V^\prime = \Lambda_\phi^4/f_\phi =\Lambda_\phi^4 g/M^2 $, we have $\alpha \simeq 0.05$ and $\beta \simeq 2\times 10^4$ so that $\alpha \beta \simeq 10^3$ (global QBC) and $\alpha^2 \beta \simeq 70$ (\QtoV marginally). Thus, the width is $\sigma_\varphi \sim \beta^{-1/2} \sim 0.7\%$ of the field domain $\phi = [ 0, \phi_*]$. This is not so small portion; actually, our benchmark was chosen to ensure this. Therefore, numerical evaluation of $\rho(\phi)$ in this case does not suffer much technical difficulties. We can convert $\rho(\phi)$ straightforwardly to $\rho(v_h)$ using the relation $v_h^2 = g\phi/\lambda_h$. In this way, Fig.~3 of the main text is obtained.

More generally, if the width $\sigma_\varphi$ turns out to be a very small portion of the field domain, then one can use the scaling behavior of \App{app:scaling} to reduce the domain and make $\sigma_\varphi$ larger. In this way, convenient field ranges can be chosen to avoid technical difficulties and readily evaluate \Eq{eq:eqsol} for equilibrium distributions.

%%%%%%%%

\end{document}